\documentclass[12pt,a4paper]{article}

\usepackage{amsmath,amssymb,amsthm,geometry}
\usepackage{mathrsfs}
\usepackage{hyperref}
\usepackage[table]{xcolor}   
\usepackage{graphicx}   
\usepackage{rotating}
\usepackage{amsmath}
\usepackage{cite}
\usepackage{longtable}
\usepackage{booktabs}

\usepackage{authblk}

\newcommand{\bb}[1]{\mathbb{#1}}

\newcommand{\Tr}{\mathrm{Tr}}
\newcommand{\hc}{\mathrm{h.c.}}

\title{Higgs and Yukawa Structure in a Clifford Algebra Model with Three Generations and $S_3$ Family Symmetry}

\author{Niels Gresnigt\\
\small{Department of Physics, Xi’an Jiaotong-Liverpool University, 111 Ren’ai Rd., Suzhou 215123, P.R. China}\\
\small{\texttt{niels.gresnigt@xjtlu.edu.cn}}}
\date{} 

\begin{document}
\maketitle

\begin{abstract}
We construct the Higgs and Yukawa sectors as a structural completion of an algebraic three-generation model based on the complex Clifford algebra $\bb{C}\ell(10)$ with an intrinsic $S_3$ family symmetry. This addresses a common limitation of algebraic frameworks, in which Standard Model fermion multiplets and gauge symmetries may be described naturally, while the Higgs and Yukawa sectors remain less developed or absent. In the present framework, three algebraically distinguished fermion sectors are permuted by $S_3$, while the Standard Model gauge generators remain generation-independent.

Higgs components are realised as right-action operators mapping weak-doublet fermion sectors into the corresponding weak-singlet sectors, and Yukawa coefficients are extracted using a Hilbert--Schmidt trace pairing. This yields two first-generation Higgs doublets with electroweak quantum numbers $(1,2,-1)$ and $(1,2,+1)$ under $SU(3)_C\times SU(2)_L\times U(1)_Y$, together with a Type-II-like separation between down-type and up-type Yukawa channels. Acting with the order-three family generator then generates a family-resolved Higgs sector organised into cyclic $S_3$ orbits.

In the cyclically averaged Higgs limit, the Type-II-like Yukawa selection rule is preserved, while the generation-space Yukawa matrix is fixed algebraically and is non-diagonal in the algebraic generation basis. Under the usual implementation of electroweak symmetry breaking, the neutral Higgs couplings are aligned with the corresponding mass matrices, so tree-level flavour-changing neutral currents are not expected in this limit. The result is a constrained algebraic starting point for future $S_3$-breaking flavour phenomenology.
\end{abstract}

\noindent\textbf{Keywords:} Higgs sector; Yukawa couplings; flavour structure; three fermion generations; $S_3$ family symmetry; Clifford algebra; Standard Model.


\section{Introduction}

Algebraic approaches to the Standard Model (SM) aim to explain fermion multiplets, gauge symmetries, and generation structure from underlying mathematical principles rather than treating them as independent inputs. A persistent limitation of many such approaches, however, is that the Higgs and Yukawa sectors remain underdeveloped or absent. This is a significant gap, because the Higgs sector is not merely additional field content: it controls electroweak symmetry breaking, fermion masses, flavour structure, and the pattern of Yukawa couplings. Thus, if an algebraic model is to move beyond a representation-theoretic account of fermions and gauge symmetries, it must also explain how Higgs degrees of freedom and Yukawa maps arise within the same algebraic framework.

In this paper we address this issue for an algebraic three-generation framework based on the complex Clifford algebra $\bb{C}\ell(10)$. The construction builds on earlier work in which three linearly independent fermion generations were realised using $\bb{C}\ell(8)$ \cite{Gillard2019,Gresnigt2023,gourlay2024algebraic}\footnote{See also \cite{gillard2019c,gresnigt2019sedenions,gresnigt2023toward,gresnigt2023sedenion}.}, together with its electroweak extension to $\bb{C}\ell(10)$ \cite{gresnigt2026electroweak}. In this framework, the discrete group $S_3$, arising from the automorphism structure of the sedenions \cite{brown1967generalized}, acts as a family symmetry on algebraic spinors. The three fermion generations are represented by algebraically distinguished, linearly independent subspaces carrying equivalent SM gauge representations, while the SM gauge generators are required to commute with the $S_3$ action. In this way the fermion content is organised into three families, but the gauge sector remains generation-independent and is not triplicated.

The aim of the present paper is to construct the Higgs and Yukawa sector inside this same algebraic framework. We realise Higgs components as right-action operators mapping weak-doublet fermion sectors into the appropriate weak-singlet sectors, and extract the corresponding Yukawa coefficients using a Hilbert--Schmidt trace pairing. This yields two first-generation Higgs doublets with the correct electroweak quantum numbers, transforming as $(1,2,-1)$ and $(1,2,+1)$ under $SU(3)_C\times SU(2)_L\times U(1)_Y$, together with a natural Type-II-like separation between down-type and up-type Yukawa channels.

Acting with the order-three family generator $\psi_3$ on the first-generation Higgs doublets generates a family-resolved Higgs sector organised into cyclic orbits. We show that the resulting Higgs states are linearly independent, and moreover linearly independent from the 96-dimensional fermion sector. Averaging over each order-three orbit defines a pair of $\psi_3$-invariant Higgs doublets. We refer to the restriction to these averaged doublets as the cyclically averaged Higgs limit. These averaged doublets are not singlets under the full $S_3$ action; rather, as shown explicitly below, they transform in the sign representation $1'$ under the order-two generator. In the cyclically averaged Higgs limit, the Type-II-like Yukawa selection rule is preserved: the down-type Higgs doublet couples only to the down-type singlet sector, while the up-type Higgs doublet couples only to the up-type singlet sector.

The generation-space Yukawa matrices are nevertheless not diagonal in the algebraic generation basis. This is because the three algebraic generation subspaces are linearly independent but not mutually orthogonal under the Hilbert--Schmidt trace pairing. In the cyclically averaged Higgs limit, the resulting Yukawa texture is fixed algebraically, with diagonal entries $\frac12$ and off-diagonal entries $\frac18$. Thus this limit distinguishes a family-symmetric direction and leaves a degenerate orthogonal pair. Under the usual implementation of electroweak symmetry breaking, the neutral Higgs couplings are aligned with the corresponding mass matrices, so tree-level flavour-changing neutral currents (FCNCs) are not expected in this limit.

This construction connects naturally with the broader phenomenology of multi-Higgs and $S_3$-based flavour models. In conventional multi-Higgs theories, the Yukawa sector generically leads to tree-level FCNCs unless special conditions are satisfied \cite{glashow1977natural,paschos1977diagonal,branco2012theory}. Likewise, realistic $S_3$ flavour models typically require a nontrivial Higgs-family structure together with some form of symmetry breaking or vacuum alignment \cite{mondragon2007lepton,teshima2012higgs,cogollo2016two,babu2024fermion}. The present framework differs conceptually from such models in that the family symmetry is algebraically motivated from the outset, and the associated Higgs-family structure is inherited from the same algebraic origin rather than imposed independently.

The cyclically averaged Higgs limit studied here should therefore be viewed as a controlled structural starting point, rather than as a complete realistic flavour model. Its main role is to establish the Higgs and Yukawa sectors of the algebraic $\bb{C}\ell(10)$ three-generation framework and to clarify the flavour structure generated by the order-three orbit average. Realistic masses, mixing angles, and CP-violating phases are expected to require departures from this cyclically averaged Higgs limit, for example through $S_3$ breaking, vacuum alignment, or vacuum expectation values in the orthogonal Higgs directions. These questions, together with the scalar potential, will be studied in future work.

A number of algebraic approaches have explored whether the existence of three fermion generations can be understood from deeper mathematical structure. These include triality-based proposals, constructions involving larger Clifford algebras, and approaches based on exceptional Jordan structures \cite{silagadze1994so,manogue1999dimensional,gording2020unified,dubois2016exceptional1,dubois2019exceptional2,todorov2018octonions,todorov2018deducing,boyle2020standard2,boyle2020standard}. Both triality-based and Clifford-algebraic considerations have pointed to $\bb{C}\ell(8)$ as a natural setting for three-generation structure \cite{silagadze1994so,gourlay2024algebraic,furey2025three,quinta2025spacetime}. The present work belongs to this broader algebraic-unification programme, but focuses on a different aspect: the internal construction of Higgs degrees of freedom and Yukawa maps once a three-generation fermion sector and an electroweak gauge action have been specified.

Some recent algebraic three-generation approaches have also touched on Higgs or mass-generation questions, but in rather different ways. In the triality-based model of \cite{furey2025three}, the Yukawa sector is discussed through Cartan factorisation and a proposed non-degenerate trilinear-form constraint. In another recent construction based on $C\ell(8,0)$, a Higgs doublet appears in the mass terms, although the bosonic sector remains to be fully developed \cite{quinta2025spacetime}. By contrast, the aim here is to construct the Higgs sector explicitly within the $\bb{C}\ell(10)$ three-generation framework itself, to identify the corresponding algebraic Yukawa maps, and to determine the resulting flavour texture in the cyclically averaged Higgs limit.

The structure of the paper is as follows. In Section 2 we summarise the algebraic ingredients of the $\bb{C}\ell(10)$ three-generation construction needed for the Higgs analysis. Section 3 constructs the Higgs components as right-action maps and introduces the trace pairing used to extract Yukawa coefficients. Section 4 studies the cyclic family orbit of the Higgs sector and constructs the corresponding cyclically averaged Higgs doublets. Section 5 analyses the resulting flavour texture in the cyclically averaged Higgs limit and discusses the associated natural-flavour-conservation condition. Section 6 summarises the main results and outlines the next steps involving $S_3$ breaking, scalar potentials, and realistic flavour phenomenology. The algebraic calculations presented in this work have been independently verified in Mathematica \cite{VerificationRepository2}.

\section{Algebraic Setup}\label{sec:algebraicsetup}

In this section we summarise the algebraic ingredients of the \(\bb{C}\ell(10)\) three-generation construction needed in what follows, and refer the reader to \cite{gresnigt2026electroweak} for additional details.

We begin by defining the following generating basis for the complex Clifford algebra $\bb{C}\ell(8)\cong \textrm{Mat}(16,\bb{C})$:
\begin{equation}
\label{Cl8basis_summary}
\begin{aligned}
e_1 &= i \sigma_1 \otimes \sigma_1 \otimes \sigma_1 \otimes \sigma_1, &\qquad
e_2 &= i \sigma_1 \otimes \sigma_1 \otimes \sigma_3 \otimes \bb{I}_2, \\
e_3 &= -i \sigma_1 \otimes \sigma_1 \otimes \sigma_1 \otimes \sigma_3, &\qquad
e_4 &= -i \sigma_1 \otimes \sigma_3 \otimes \bb{I}_2 \otimes \bb{I}_2, \\
e_5 &= -i \sigma_1 \otimes \sigma_1 \otimes \sigma_1 \otimes \sigma_2, &\qquad
e_6 &= -i \sigma_1 \otimes \sigma_2 \otimes \bb{I}_2 \otimes \bb{I}_2, \\
e_7 &= i \sigma_1 \otimes \sigma_1 \otimes \sigma_2 \otimes \bb{I}_2, &\qquad
e_8 &= i \sigma_2 \otimes \bb{I}_2 \otimes \bb{I}_2 \otimes \bb{I}_2,
\end{aligned}
\end{equation}
with the $\bb{C}\ell(8)$ pseudoscalar defined as $\omega_8:=e_1e_2e_3e_4e_5e_6e_7e_8$.

Subsequently, define a Witt basis for the algebra as:
\begin{eqnarray}
a_j=\frac{1}{2}(-e_j+i e_{j+4}),\qquad a_j^\dagger=\frac{1}{2}(e_j+i e_{j+4}),\qquad j=1,\ldots,4.
\end{eqnarray}
These Witt basis generators satisfy the fermionic oscillator relations
\begin{eqnarray}
\{a_i,a_j\}=0,\qquad \{a_i^\dagger,a_j^\dagger\}=0,\qquad \{a_i,a_j^\dagger\}=\delta_{ij}.
\end{eqnarray}
For each Witt pair we define two complementary simple idempotents, defined as
\begin{eqnarray}
\pi_j^{(+)}=a_j a_j^\dagger,\quad \pi_j^{(-)}=a_j^\dagger a_j,\quad \pi_j^{(\pm)}\pi_k^{(\pm)}=\pi_k^{(\pm)}\pi_j^{(\pm)}\quad j=1,\ldots,4.
\end{eqnarray}
From these, the primitive idempotents are given by
\begin{eqnarray}
f_{\varepsilon_1\varepsilon_2\varepsilon_3\varepsilon_4}
=
\pi_1^{(\varepsilon_1)}\pi_2^{(\varepsilon_2)}\pi_3^{(\varepsilon_3)}\pi_4^{(\varepsilon_4)},
\qquad \varepsilon_i\in\{+,-\}.
\end{eqnarray}
These primitive idempotents are mutually orthogonal and complete:
\begin{equation}
f_{\varepsilon} f_{\varepsilon'} = \delta_{\varepsilon_1,\varepsilon'_1}\delta_{\varepsilon_2,\varepsilon'_2}\delta_{\varepsilon_3,\varepsilon'_3}\delta_{\varepsilon_4,\varepsilon'_4} f_{\varepsilon},
\qquad
\sum_{\varepsilon\in\{+,-\}^4} f_{\varepsilon} = \bb{I}_{16},\qquad f_\varepsilon=f_{\varepsilon_1\varepsilon_2\varepsilon_3\varepsilon_4}.
\end{equation}

We denote by $E_{ij}$ the standard matrix unit with a single nonzero entry equal to $1$ in row $i$ and column $j$. In particular,
\begin{eqnarray}
f_{++++}=E_{1,1},\qquad f_{---+}=E_{14,14},\qquad f_{----}=E_{6,6},\qquad f_{+++-}=E_{9,9}.
\end{eqnarray}
The primitive idempotents are used to decompose the algebra $\bb{C}\ell(8)$ into its minimal left ideals. In the chosen matrix realisation, with $f_{++++}=E_{1,1}$, the first minimal left ideal is
\begin{eqnarray}
I_1=\bb{C}\ell(8)f_{++++}
=\textrm{Mat}(16,\bb{C})\,E_{1,1}
=\mathrm{span}_{\bb{C}}\{E_{j,1}\}_{j=1}^{16},
\end{eqnarray}
corresponding to the subspace of matrices with support only in the first column.

The family symmetry that organises the three fermion families is generated by the order-three map $\psi_3$ and the order-two map $\epsilon$. The action of $\psi_3$ on the Clifford generators is given by
\begin{eqnarray}
\psi_3: e_i&\mapsto& \frac{1}{4}e_i+\frac{\sqrt{3}}{4}g_i-\frac{\sqrt{3}}{4}e_i e_8-\frac{3}{4}g_i e_8,\qquad i=1,\ldots,7,\\
\psi_3: e_8&\mapsto& e_8,
\end{eqnarray}
where
\begin{eqnarray}
\nonumber g_1 &=& \tfrac{1}{2}e_1e_8(-e_{2345}+e_{2367}+e_{4567}-1) ,\quad g_{2} = \tfrac{1}{2}e_2e_8(e_{1346}+e_{1357}+e_{4567}-1), \\ 
g_{3} &=& \tfrac{1}{2}e_3e_8(e_{1256}-e_{1247}+e_{4567}-1), \quad g_{4} =\tfrac{1}{2}e_4e_8(e_{1256}+e_{1357}+e_{2367}-1), \\ 
\nonumber g_{5} &=& \tfrac{1}{2}e_5e_8(-e_{1247}+e_{1346}+e_{2367}-1), \quad g_{6} = \tfrac{1}{2}e_6e_8(-e_{1247}+e_{1357}-e_{2345}-1), \\ 
\nonumber g_{7} &=& \tfrac{1}{2}e_7e_8(e_{1256}+e_{1346}-e_{2345}-1),
\end{eqnarray}
and $e_{i_1i_2\cdots i_k}:=e_{i_1}e_{i_2}\cdots e_{i_k}$.

The order-two generator meanwhile is defined as
\begin{eqnarray}
\epsilon:e_i\mapsto -e_i,\qquad i=1,\ldots,8.
\end{eqnarray}
Together, $\psi_3$ and $\epsilon$ satisfy $\psi_3^3=\mathrm{Id},\quad \epsilon^2=\mathrm{Id},\quad \epsilon\psi_3=\psi_3^2\epsilon$, and generate $S_3$.

\subsection{$\bb{C}\ell(10)$ extension and $S_3$ action}

Subsequently, the algebra $\bb{C}\ell(8)$ is enlarged to $\bb{C}\ell(10)$ via the graded tensor product embedding
\begin{eqnarray}
\iota:\bb{C}\ell(8)\hookrightarrow \bb{C}\ell(10),\qquad \iota(e_i)=\bb{I}_2\otimes e_i,\qquad i=1,\ldots,8,
\end{eqnarray}
with the remaining two basis elements defined by
\begin{eqnarray}
e_9:=-i\bar{\sigma}_1\otimes \omega_8,\qquad e_{10}:=i\bar{\sigma}_2\otimes \omega_8,
\end{eqnarray}
where $\bar{\sigma}_i$ are Pauli matrices, and the bar is used to avoid confusion with the Pauli matrices used to define our $\bb{C}\ell(8)$ basis. The fifth Witt pair is then
\begin{eqnarray}
a_5=\frac{1}{2}(-e_9+i e_{10}),\qquad a_5^\dagger=\frac{1}{2}(e_9+i e_{10}),
\end{eqnarray}
with associated simple idempotents
\begin{eqnarray}
\pi_5^{(+)}=a_5 a_5^\dagger,\qquad \pi_5^{(-)}=a_5^\dagger a_5.
\end{eqnarray}
Accordingly, the primitive idempotents of $\bb{C}\ell(10)$ are
\begin{eqnarray}
f_{\varepsilon_1\varepsilon_2\varepsilon_3\varepsilon_4\varepsilon_5}
=
\pi_1^{(\varepsilon_1)}\pi_2^{(\varepsilon_2)}\pi_3^{(\varepsilon_3)}\pi_4^{(\varepsilon_4)}\pi_5^{(\varepsilon_5)}, \qquad \varepsilon_i\in\{+,-\}.
\end{eqnarray}
In the corresponding $32\times 32$ matrix realisation,
\begin{eqnarray}
f_{+++++}=E_{1,1},\qquad f_{++++-}=E_{17,17},\qquad f_{+++-+}=E_{9,9},\qquad f_{+++--}=E_{25,25}.
\end{eqnarray}

The action of the $S_3$ generators on $\bb{C}\ell(8)$ extends naturally to $\bb{C}\ell(10)$ through the graded tensor product decomposition
\begin{eqnarray}
\bb{C}\ell(10)\cong \bb{C}\ell(2)\,\hat{\otimes}\,\bb{C}\ell(8).
\end{eqnarray}
The action of $S_3$ on the $\bb{C}\ell(2)$ factor is taken to be trivial, so in particular $\psi_3(\bar{\sigma}_i)=\bar{\sigma}_i$ and $\epsilon(\bar{\sigma}_i)=\bar{\sigma}_i$ for $i=1,2,3$. On the embedded subalgebra $\iota(\bb{C}\ell(8))=\bb{I}_2\otimes\bb{C}\ell(8)$, the maps $\psi_3$ and $\epsilon$ therefore act block-diagonally on the two $\bb{C}\ell(8)$ blocks, so that
\begin{eqnarray}
\iota:\psi_3\mapsto \psi_3\oplus\psi_3,\qquad \iota:\epsilon\mapsto \epsilon\oplus\epsilon.
\end{eqnarray}
Since the additional generators $e_9$ and $e_{10}$ are defined using the $\bb{C}\ell(8)$ pseudoscalar $\omega_8$, their transformation is induced from the action of $S_3$ on $\omega_8$. Thus $\psi_3$ acts nontrivially on $e_9$ and $e_{10}$, whereas $\epsilon(\omega_8)=\omega_8$ and hence fixes the fifth Witt pair:
\begin{eqnarray}
\epsilon(a_5)=a_5,\qquad \epsilon(a_5^\dagger)=a_5^\dagger.
\end{eqnarray}
It follows that the three Witt bases
\begin{eqnarray}
\{a_i,a_i^\dagger\},\qquad \{\psi_3(a_i),\psi_3(a_i^\dagger)\},\qquad \{\psi_3^2(a_i),\psi_3^2(a_i^\dagger)\},\qquad i=1,\ldots,5,
\end{eqnarray}
again satisfy the fermionic anticommutation relations. In this way the $S_3$ family symmetry of $\bb{C}\ell(8)$ lifts consistently to $\bb{C}\ell(10)$, and continues to permute the relevant primitive idempotents and fermion subspaces without replicating the gauge generators.

\subsection{Fermion subspaces and three generations}
In the $32\times 32$ matrix realisation, we write
\begin{eqnarray}
I_m:=\bb{C}\ell(10)E_{m,m},
\end{eqnarray}
so that $I_m$ denotes the minimal left ideal supported on the $m$th column.

The first-generation isospin-up doublet sector is
\begin{equation}
V_1^+:=\Bigl(\mathrm{span}_{\bb{C}}\bigl\{
a_1^\dagger a_2^\dagger a_3^\dagger a_4^\dagger,\;
a_i^\dagger a_j^\dagger,\;
a_i^\dagger a_4^\dagger,\;
a_4^\dagger a_5^\dagger
\bigr\}\Bigr)f_{+++++}\subset I_1,
\label{eq:V1plus_full_again}
\end{equation}
with \(i,j\in\{1,2,3\}\) and \(i<j\). The corresponding isospin-down doublet sector is
\begin{equation}
V^{-}_{1}:=
\left(
\mathrm{span}_{\bb{C}}
\left\{
a_1^\dagger a_2^\dagger a_3^\dagger a_4^\dagger a_5,\,
a_i^\dagger a_j^\dagger a_5,\,
a_i^\dagger a_4^\dagger a_5,\,
a_4^\dagger
\right\}
\right)f_{++++-}\subset I_{17}.
\label{eq:V1minus_full_again}
\end{equation}

The weak singlet sectors are
\begin{equation}
U^{-}_{1}:=
\left(
\mathrm{span}_{\bb{C}}
\left\{
a_1^\dagger a_2^\dagger a_3^\dagger,\,
a_i^\dagger a_j^\dagger a_4,\,
a_i^\dagger,\,
a_5^\dagger
\right\}
\right)f_{+++-+}\subset I_9,
\label{eq:U1minus_again}
\end{equation}
and
\begin{equation}
U^{+}_{1}:=
\left(
\mathrm{span}_{\bb{C}}
\left\{
a_1^\dagger a_2^\dagger a_3^\dagger a_5,\,
a_i^\dagger a_j^\dagger a_4 a_5,\,
a_i^\dagger a_5,\,
1
\right\}
\right)f_{+++--}\subset I_{25}.
\label{eq:U1plus_again}
\end{equation}

These fermion subspaces are related by right multiplication:
\begin{eqnarray}
    V_1^-=V_1^+a_5,\qquad U_1^-=V_1^+a_4,\qquad U_1^+=V_1^+a_4a_5.
\end{eqnarray}
Here these equalities are equalities of subspaces. Right multiplication by $a_5$ and $a_4$ changes the relevant oscillator occupation, and therefore maps each algebraic representative to the corresponding state in the target weak-doublet or weak-singlet sector. In particular, the $a_5$ map includes the expected relabelling of weak partners, for example $u_L\mapsto d_L$ and $\nu_L\mapsto e^-_L$, up to signs fixed by the chosen ordering of the Witt generators.

Equivalently,
\begin{eqnarray}
    U_1^-=V_1^-a_4a_5^{\dagger},\qquad U_1^+=V_1^-a_4.
\end{eqnarray}

The first-generation fermion subspace is then
\begin{eqnarray}
S_1:=V_1^+\oplus V_1^-\oplus U_1^-\oplus U_1^+.
\end{eqnarray}
with
\begin{eqnarray}
V_1^+ &\leftrightarrow& \left(e_L^+,\;\bar{d}_L^{(\bar{3})},\;u_L^{(3)},\;\nu_L\right),\quad V_1^- \leftrightarrow \left(\bar{\nu}_L,\;\bar{u}_L^{(\bar{3})},\;d_L^{(3)},\;e_L^-\right),\\
U_1^+ &\leftrightarrow& \left(e_R^+,\;\bar{d}_R^{(\bar{3})},\;u_R^{(3)},\;\nu_R\right),\quad U_1^- \leftrightarrow \left(\bar{\nu}_R,\;\bar{u}_R^{(\bar{3})},\;d_R^{(3)},\;e_R^-\right).
\end{eqnarray}
Here an overbar denotes the corresponding antiparticle state, so for example $\bar{u}$ and $\bar{d}$ denote antiup and antidown quarks, respectively. The explicit algebraic representation of the first generation of fermion states is given in Appendix \ref{sec:appA}.

The remaining two generations are obtained by the order-three family action: $S_2:=\psi_3(S_1),\; S_3:=\psi_3^2(S_1)$.
Thus $\psi_3$ cyclically permutes the three generation subspaces. These subspaces are linearly independent and carry equivalent SM gauge representations.

\subsection{Gauge generators and their action}
The $SU(3)_C$ colour gauge generators in $\bb{C}\ell(10)$ are given by
\begin{eqnarray}
\nonumber\Lambda_1 &=&  -a_2^\dagger{a_1}-a_1^\dagger{a_2}, \quad 
\Lambda_2 = ia_2^\dagger{a_1}-ia_1^\dagger{a_2}, \quad \Lambda_3 = a_2^\dagger{a_2}-a_1^\dagger{a_1}, \\
\Lambda_4 &=& -a_1^\dagger{a_3}-a_3^\dagger{a_1}, \quad 
\Lambda_5 = -ia_1^\dagger{a_3}+ia_3^\dagger{a_1}, \quad
\Lambda_6 = -a_3^\dagger{a_2}-a_2^\dagger{a_3}, \\ 
\nonumber\Lambda_7 &=& ia_3^\dagger{a_2}-ia_2^\dagger{a_3}, \quad
\Lambda_8 = -\frac{1}{\sqrt{3}}(a_1^\dagger{a_1}+a_2^\dagger{a_2}-2a_3^\dagger{a_3}),
\end{eqnarray}

The $SU(2)_L$ weak gauge generators are defined as
\begin{eqnarray}\label{eq:weakgenerators}
T_1&=&\frac{1}{2}(-i a_5+i a_5^\dagger)\omega_8 P,\quad T_2=\frac{1}{2}(a_5+a_5^\dagger)\omega_8 P,\quad T_3=\frac{1}{2}(a_5^\dagger a_5-a_5 a_5^\dagger)P,
\end{eqnarray}
where $P$ is an $S_3$-invariant projector defined as
\begin{eqnarray}
P:=f_{+++++}+f_{---++}+f_{++++-}+f_{---+-}.
\end{eqnarray}

The first-generation $SU(2)_L$ doublets are then
\begin{eqnarray}
\left(\begin{array}{c}
a_1^\dagger a_2^\dagger a_3^\dagger a_4^\dagger f_{+++++}\\
a_1^\dagger a_2^\dagger a_3^\dagger a_4^\dagger a_5 f_{++++-}
\end{array}
\right)&,&\;
\left(\begin{array}{c}
a_i^\dagger a_j^\dagger f_{+++++}\\
a_i^\dagger a_j^\dagger a_5 f_{++++-}
\end{array}
\right),\;\\
\left(\begin{array}{c}
a_i^\dagger a_4^\dagger f_{+++++}\\
a_i^\dagger a_4^\dagger a_5 f_{++++-}
\end{array}
\right)&,&\;
\left(\begin{array}{c}
a_4^\dagger a_5^\dagger f_{+++++}\\
a_4^\dagger f_{++++-}
\end{array}
\right),
\end{eqnarray}
where again $i,j\in\{1,2,3\}$ with $i<j$. 

The weak-isospin eigenvalues are defined with respect to the commutator action. Although the right action of $T_3$ alone gives the opposite sign, on the doublet subspaces $V_1^\pm$ one has $[T_i,X]=-XT_i$. Hence the states in $V_1^+$ have physical weak isospin $T_3=+1/2$, while those in $V_1^-$ have $T_3=-1/2$.

The $U(1)_{em}$ electromagnetic charge generator is defined as
\begin{eqnarray}\label{eq:Q'def}
Q'&:=&Q+(2P-\bb{I})a_5^\dagger a_5,
\end{eqnarray}
where
\begin{eqnarray}\label{eq:Qdef}
Q&:=&\frac{1}{3}\left(Q_1+\psi_3(Q_1)+\psi_3^2(Q_1)\right),
\end{eqnarray}
is the $\psi_3$-invariant charge operator obtained by averaging the first-generation generator
\begin{eqnarray}\label{eq:Q1def}
Q_1&:=&\frac{1}{3}\left(a_1a_1^\dagger+a_2a_2^\dagger+a_3a_3^\dagger-3a_4a_4^\dagger\right),
\end{eqnarray}
over its $\psi_3$ orbit.

Finally, the $U(1)_Y$ hypercharge generator is given by
\begin{eqnarray}
   Y&:=&2(Q'-T_3). 
\end{eqnarray}

The gauge generators act on the fermion states via the adjoint (commutator) action $[.,.]$. On the physically relevant fermion subspaces this action simplifies: the colour generators $\Lambda_i$ reduce to a left action, while the weak generators $T_i$ reduce to a right action, with the sign convention described above. This is precisely the structural feature that makes it natural to introduce the Higgs below as a right-action operator from the weak-doublet subspaces $V_1^\pm$ to the weak-singlet subspaces $U_1^\pm$.

A crucial property for the present construction is that the SM gauge generators are $S_3$-invariant. In particular, the action of the family symmetry permutes the fermion subspaces while leaving the gauge algebra fixed, so the gauge sector is generation-independent and is not triplicated.

\section{The Higgs Sector}

In this section we develop the Higgs sector in $\bb{C}\ell(10)$. We treat the Higgs components as right-action operators that map the weak-doublet subspaces into the appropriate weak-singlet subspaces. An algebraic Yukawa coefficient can then be extracted using a suitable bilinear pairing. For the latter we use the Hilbert--Schmidt inner product in the matrix realisation $\bb{C}\ell(10)\cong\text{Mat}(32,\bb{C})$.

\subsection{Why a direct trilinear product is insufficient}

At first glance one might try to proceed exactly as in ordinary field theory and simply replace the spacetime fields in a term such as
\begin{eqnarray}
\overline{\psi}_L\, H\, \psi_R
\end{eqnarray}
by the corresponding algebraic representatives from the $\bb{C}\ell(10)$ model\footnote{Here \(\overline{\psi}_L\) denotes the usual Dirac adjoint of the spacetime spinor field,
and should not be confused with the overbar used in this paper to label
antiparticle states such as \(\bar{u}\), \(\bar{d}\), or \(\bar{\nu}\).}. One would then attempt to determine the Higgs components by demanding that the resulting algebraic product be nonzero. This idea is natural, but it is not the correct algebraic notion of a Yukawa coupling. For example, consider an attempted colour-diagonal down-type channel. From Table~\ref{tab:first-generation-states}, the internal algebraic state carrying the conjugate gauge quantum numbers of a left-handed down quark is represented, for a fixed colour choice, by
\begin{eqnarray}
\bar d_L = a_2^\dagger a_3^\dagger f_{+++++},\qquad
d_R = a_1^\dagger f_{+++-+}.
\end{eqnarray}
A naive attempt to mimic the spacetime Yukawa factor $\bar d_L H d_R$ by replacing the barred spacetime spinor with this internal antiparticle-labelled algebraic state would then lead one to consider
\begin{eqnarray}
a_2^\dagger a_3^\dagger
\left(f_{+++++}Hf_{-++-+}\right)
a_1^\dagger \neq 0,
\end{eqnarray}
where we have used $a_1^\dagger f_{+++-+}=f_{-++-+}a_1^\dagger$. Demanding that the term in parentheses be nonzero forces the Higgs component to contain a colour-carrying factor such as
\begin{eqnarray}
H\sim a_1a_4f_{-++-+},
\end{eqnarray}
which is immediately unacceptable as a Higgs candidate.

The reason this simplistic approach fails is that the algebraic product of three state representatives is not the correct analogue of a spacetime Yukawa invariant. In ordinary field theory, the barred spinor $\bar{\psi}$ is not simply another particle state but instead a Dirac adjoint spinor that belongs to the dual space. Therefore one should not expect the raw algebraic product of three state representatives to be the final physical
invariant.

\subsection{Higgs operators as right-action maps}
Within the $\bb{C}\ell(10)$ framework, the weak-doublet states lie in the subspaces $V_1^+$ and $V_1^-$, while the weak-singlet states lie in the subspaces $U_1^+$ and $U_1^-$. These subspaces belong to distinct minimal left ideals. The Yukawa terms therefore connect states belonging to different minimal-ideal sectors, suggesting that the Higgs components should be treated as right-action operators mapping weak-doublet subspaces into weak-singlet subspaces. A suitable bilinear pairing can subsequently be used to extract an algebraic Yukawa coefficient.

Let us first identify Higgs operators \(H\) for which the map
\begin{equation}
R_H : X_L \mapsto X_L H
\label{eq:right_action_intertwiner_preview}
\end{equation}
takes weak doublet states into weak singlet states. Here \(X_L\in \bb{C}\ell(10)\) is an
internal algebraic state representing a left-handed fermion.

Already at the algebraic level, the two singlet sectors suggest the relevant Higgs maps. The maps
\begin{eqnarray}
V_1^+ \xrightarrow{\ a_4\ } U_1^-,
\quad
V_1^- \xrightarrow{\ a_4 a_5^\dagger\ } U_1^-,
\quad
V_1^+ \xrightarrow{\ a_4 a_5\ } U_1^+,
\qquad
V_1^- \xrightarrow{\ a_4\ } U_1^+,
\end{eqnarray}
are precisely the ones needed to reproduce the Yukawa patterns. This suggests that the Higgs sector should consist of two weak doublets: one down-type doublet mapping to \(U_1^-\), and one up-type doublet mapping to \(U_1^+\).

The first-generation fermion states of the \(\bb{C}\ell(10)\) model are organised into the weak
doublet sectors
\begin{equation}
V_1^+ \subset I_1 = \bb{C}\ell(10)f_{+++++},
\qquad
V_1^-  \subset I_{17} = \bb{C}\ell(10)f_{++++-},
\label{eq:V1pm_again}
\end{equation}
and the weak singlet sectors
\begin{equation}
U_1^- \subset I_9 = \bb{C}\ell(10)f_{+++-+},
\qquad
U_1^+  \subset I_{25} = \bb{C}\ell(10)f_{+++--}.
\label{eq:U1pm_again}
\end{equation}
The target primitive idempotents are therefore
\begin{equation}
f_{+++-+}=E_{9,9},
\qquad
f_{+++--}=E_{25,25},
\label{eq:target_idempotents_again}
\end{equation}
whereas the source primitive idempotents for the weak doublets are
\begin{equation}
f_{+++++}=E_{1,1},
\qquad
f_{++++-}=E_{17,17}.
\label{eq:source_idempotents_again}
\end{equation}

For primitive idempotents \(f_a\) and \(f_b\), define
\begin{equation}
H_{a\to b} := f_a\,\bb{C}\ell(10)\,f_b.
\label{eq:intertwiner_space_def}
\end{equation}

For the physical Higgs maps of interest there are four relevant operator spaces:
\begin{equation}
H_{1\to 9} := f_{+++++}\bb{C}\ell(10)f_{+++-+},
\qquad
H_{17\to 9} := f_{++++-}\bb{C}\ell(10)f_{+++-+},
\label{eq:Hd_intertwiner_spaces}
\end{equation}
\begin{equation}
H_{1\to 25} := f_{+++++}\bb{C}\ell(10)f_{+++--},
\qquad
H_{17\to 25} := f_{++++-}\bb{C}\ell(10)f_{+++--}.
\label{eq:Hu_intertwiner_spaces}
\end{equation}

Each such space is one-dimensional, so once the source and target primitive idempotents are fixed, the corresponding Higgs operator is unique up to an overall scalar. A convenient nonzero generator of $H_{1\to 9}$ is
\begin{eqnarray}
H_{1\to 9} &=& f_{+++++}\bb{C}\ell(10)f_{+++-+},\\
&=& f_{+++++}a_4f_{+++-+},\\
&=& a_4f_{+++-+}f_{+++-+},
\qquad (\text{since } f_{+++++}a_4=a_4f_{+++-+}),\\
&=& a_4f_{+++-+},
\qquad (\text{since } f_{+++-+}f_{+++-+}=f_{+++-+}).
\end{eqnarray}
Likewise,
\begin{eqnarray}
H_{17\to 9} = a_4a_5^{\dagger}f_{+++-+},
\qquad
H_{1\to 25} = a_4a_5f_{+++--},
\qquad
H_{17\to 25} = a_4f_{+++--}.
\end{eqnarray}
In the chosen matrix realisation, each space \(H_{a\to b}\) is spanned by the corresponding
matrix unit \(E_{a,b}\). By a slight abuse of notation, we use the same symbol \(H_{a\to b}\) for a chosen nonzero
generator of the one-dimensional space \(f_a\bb{C}\ell(10)f_b\).

If \(X_L\in V_1^+\), then right multiplication by \(H_{1\to 9}=a_4f_{+++-+}\) gives
\begin{equation}
R_{H_{1\to 9}}:X_L \mapsto X_L\,H_{1\to 9}
=X_L\,a_4 f_{+++-+}\in U_1^-.
\label{eq:right_map_Vplus_to_Uminus}
\end{equation}

Likewise, if $Y_L=X_La_5\in V^-_1$ with $X_L\in V^+_1$, then right multiplication by $H_{17\to 9}=a_4a_5^\dagger f_{+++-+}$ gives 
\begin{eqnarray} 
R_{H_{17\to 9}}:Y_L\mapsto Y_LH_{17\to 9} &=&X_La_5a_4a_5^\dagger f_{+++-+} \nonumber\\ 
&=&-X_La_4f_{+++-+}\in U^-_1. 
\end{eqnarray} 
The overall sign is fixed by the chosen ordering of the Witt generators and has no effect on the induced subspace map.

Similarly, right multiplication by \(H_{1\to 25}\) and \(H_{17\to 25}\) maps
\(V_1^+\) and \(V_1^-\) into \(U_1^+\), respectively. Thus the first-generation Higgs sector should be thought of as two pairs of right-action maps.

\subsection{Candidate Higgs doublets}

It is now natural to assemble the four right-action operators into two candidate
Higgs doublets.
\begin{equation}
H_d^{(1)} :=
\begin{pmatrix}
H_{17\to 9} \\
H_{1\to 9}
\end{pmatrix},\quad 
H_u^{(1)} :=
\begin{pmatrix}
H_{17\to 25} \\
H_{1\to 25}
\end{pmatrix}.
\end{equation}
We therefore identify the four right-action operators with the Higgs components as
\begin{equation}
H_{17\to 9}=H_d^0,\quad H_{1\to 9}=H_d^-,\quad H_{17\to 25}=H_u^+,\quad H_{1\to 25}=H_u^0,
\label{eq:H_component_identification}
\end{equation}
so that
\begin{equation}
H_d^{(1)} =\begin{pmatrix}
H_d^0 \\
H_d^-
\end{pmatrix}
=
\begin{pmatrix}
a_4 a_5^\dagger f_{+++-+} \\
a_4 f_{+++-+}
\end{pmatrix},\quad
H_u^{(1)} =\begin{pmatrix}
H_u^+ \\
H_u^0
\end{pmatrix}
=\begin{pmatrix}
a_4 f_{+++--} \\
a_4 a_5 f_{+++--}
\end{pmatrix}.
\end{equation}
In this section the generation superscript on the individual Higgs components is suppressed; it will be included when the $\psi_3$-images are introduced.

In the explicit \(32\times 32\) matrix representation of \(\bb{C}\ell(10)\), these Higgs
components are
\begin{equation}
H_d^0 = -E_{17,9},\qquad
H_d^- = iE_{1,9},\qquad
H_u^+ = iE_{17,25},\qquad
H_u^0 = E_{1,25}.
\label{eq:Higgs_matrix_components}
\end{equation}

\subsection{Trace pairing and spacetime Yukawa terms}

In the chosen matrix realisation \(\bb{C}\ell(10)\cong \mathrm{Mat}(32,\bb{C})\), we equip \(\bb{C}\ell(10)\) with the Hermitian pairing induced from the matrix algebra, namely the Hilbert--Schmidt inner product
\begin{equation}
\langle A,B\rangle := \Tr(A^\dagger B), \qquad A,B\in \bb{C}\ell(10).
\label{eq:trace_inner_product}
\end{equation}
In the matrix-unit basis this satisfies
\begin{equation}
\Tr(E_{ij}^\dagger E_{kl})=
\Tr(E_{ji}E_{kl})=\delta_{ik}\delta_{jl}.
\label{eq:matrix_unit_trace_pairing}
\end{equation}
So distinct matrix units are orthonormal.

This pairing allows one to extract a complex number from the image \(X_LH\) by comparing it with a chosen right-handed algebraic state \(X_R\). We therefore define the Yukawa coefficient by
\begin{equation}
\mathcal{Y}(X_L,H,X_R):=
\Tr\bigl((X_LH)^\dagger X_R\bigr).
\label{eq:Yukawa_pairing_def}
\end{equation}

Comparing \eqref{eq:Yukawa_pairing_def} with \eqref{eq:matrix_unit_trace_pairing}, we see that \(\mathcal{Y}(X_L,H,X_R)\) simply extracts the coefficient of \(X_R\) in the expansion of \(X_LH\) in the matrix-unit basis. In particular, \(\mathcal{Y}(X_L,H,X_R)=0\) whenever \(X_LH\) has no component along \(X_R\), while \(\mathcal{Y}(X_L,H,X_R)\neq 0\) whenever the \(X_R\) component is nonzero. For normalized matrix-unit basis states, \(\mathcal{Y}(X_L,H,X_R)=\pm 1\) when \(X_LH=\pm X_R\). Thus the algebra determines the selection rules and the structure of the Yukawa sector, while the overall coupling constants remain free parameters, just as in ordinary field theory.

It is important to distinguish the internal algebraic states from the physical spacetime
spinor fields. An algebraic state \(X\in\bb{C}\ell(10)\) is not itself a Dirac or Weyl spinor
in spacetime. Rather, a physical fermion field is taken to be of the form
\begin{equation}
\Psi_{L,R}(x)=\chi_{L,R}(x)\otimes X_{L,R},
\end{equation}
where \(\chi_{L,R}(x)\) is the usual spacetime spinor field and \(X_{L,R}\in\bb{C}\ell(10)\)
is the corresponding algebraic representative.

The barred spacetime field is then
\begin{equation}
\overline{\Psi}_{L,R}(x)=\overline{\chi}_{L,R}(x)\otimes X_{L,R}^\dagger,
\end{equation}
where the bar on \(\chi\) denotes the usual Dirac adjoint in spacetime, while the dagger
on \(X\) denotes Hermitian conjugation in the matrix algebra.

Once the algebraic Yukawa coefficient has been extracted, the corresponding spacetime
Yukawa term takes the form
\begin{equation}
\mathcal{L}_Y \supset
y\,\mathcal{Y}(X_L,H,X_R)\,\overline{\chi}_L(x)\,\varphi_H(x)\,\chi_R(x)+\hc,
\end{equation}
where \(y\) is an overall coupling constant, \(\varphi_H(x)\) is the spacetime Higgs field
associated with the algebraic right-action operator \(H\), and \(\mathcal{Y}(X_L,H,X_R)\)
is the algebraically computed coefficient.

\subsection{Illustrative Yukawa channels}

Before proceeding, it is useful to work through two explicit examples. 

Take the first-generation left-handed up-quark state and right-handed down-quark state
\begin{equation}
u_L^{(i)} = a_i^\dagger a_4^\dagger f_{+++++},\quad d_R^{(j)} = a_j^\dagger f_{+++-+}.
\label{eq:uL_example}
\end{equation}
Here \(i,j=1,2,3\) denote the colour labels of \(u_L\) and \(d_R\). Acting on \(u_L^{(i)}\) with the right-action operator \(H_d^-=a_4f_{+++-+}\), we obtain
\begin{equation}
R_{H_d^-}:u_L^{(i)} \mapsto u_L^{(i)} H_d^-=a_i^\dagger a_4^\dagger f_{+++++} a_4 f_{+++-+}.
\end{equation}
Using \(f_{+++++}a_4 = a_4f_{+++-+}\), this becomes
\begin{equation}
u_L^{(i)} H_d^-=a_i^\dagger a_4^\dagger a_4 f_{+++-+}.
\end{equation}
Since \(a_4^\dagger a_4 = \pi_4^{(-)}\) and \(f_{+++-+}\) already contains \(\pi_4^{(-)}\), we obtain
\begin{equation}
u_L^{(i)} H_d^-=a_i^\dagger f_{+++-+}=d_R^{(i)}.
\end{equation}
Therefore
\begin{equation}
\mathcal{Y}\bigl(u_L^{(i)},H_d^-,d_R^{(j)}\bigr)=\Tr\bigl((d_R^{(i)})^\dagger d_R^{(j)}\bigr)= \delta_{ij},
\end{equation}
so colour diagonality is enforced automatically by the trace pairing.

As a second example, consider the neutrino channel in the up-type Higgs sector. Take
\begin{equation}
\nu_L = a_4^\dagger a_5^\dagger f_{+++++},\qquad\nu_R = f_{+++--}.
\end{equation}
Acting with \(H_u^0=a_4 a_5 f_{+++--}\), we obtain
\begin{eqnarray}
R_{H_u^0}:\nu_L \mapsto\nu_L H_u^0&=&a_4^\dagger a_5^\dagger f_{+++++} a_4 a_5 f_{+++--}\nonumber\\
&=&a_4^\dagger a_5^\dagger a_4 a_5 f_{+++--}\qquad\text{using }f_{+++++}a_4a_5=a_4a_5f_{+++--}\nonumber\\
&=&-\,a_5^\dagger a_4^\dagger a_4 a_5 f_{+++--}\nonumber\\
&=&-\,\pi_5^{(-)} \pi_4^{(-)} f_{+++--}\nonumber\\
&=&-\,f_{+++--}\nonumber\\
&=&-\,\nu_R.
\end{eqnarray}
Thus the map again lands on the desired right-handed singlet, up to a sign. Subsequently
\begin{equation}
\mathcal{Y}\bigl(\nu_L,H_u^0,\nu_R\bigr)=\Tr \bigl(-\nu_R^\dagger \nu_R\bigr)= -1.
\end{equation}

Likewise, for the lower weak partner \(e_L^- = a_4^\dagger f_{++++-}\) one finds
\begin{eqnarray}
R_{H_u^+}:e_L^- \mapsto e_L^- H_u^+
&=&a_4^\dagger f_{++++-} a_4 f_{+++--}\nonumber\\
&=&\nu_R.
\end{eqnarray}
and therefore,
\begin{equation}
\mathcal{Y}\bigl(e_L^-,H_u^+,\nu_R\bigr)=\Tr\bigl(\nu_R^\dagger \nu_R\bigr)= +1.
\end{equation}
So the two members of the weak doublet map into the same right-handed singlet sector, as required.

\subsection{Gauge quantum numbers and Type-II structure}
The Higgs components are first seen to be \(SU(3)_C\) singlets. Indeed, the colour generators
\(\Lambda_i\) act only on the first three Witt modes \(a_1,a_2,a_3\), whereas the nontrivial Higgs factors are built from \(a_4\) and \(a_5\), with the remaining factor \(\pi_1^{(+)}\pi_2^{(+)}\pi_3^{(+)}\) in the primitive idempotents being colour invariant. Hence
\begin{equation}
[\Lambda_i,H]=0,\qquad i=1,\ldots,8,
\end{equation}
for \(H\in\{H_d^0,H_d^-,H_u^+,H_u^0\}\).

Using the $SU(2)_L$ generators given in \eqref{eq:weakgenerators}, one readily verifies that the two Higgs pairs $H_d$ and $H_u$ are genuine weak doublets. In particular we find that
\begin{eqnarray}
[T_3,H_d^0]=+\frac12 H_d^0,\quad [T_3,H_d^-]=-\frac12 H_d^-,\quad [T_3,H_u^+]=+\frac12 H_u^+,\quad [T_3,H_u^0]=-\frac12 H_u^0.
\end{eqnarray}

Unlike for fermions, for which the commutator action of \(SU(2)_L\) reduces to a right
action on the weak-doublet subspaces \(V_1^+\) and \(V_1^-\),
\begin{equation}
[T_i,X]=-XT_i,\qquad X\in V_1^+\oplus V_1^-,\qquad i=1,2,3,
\end{equation}
the Higgs sector behaves differently under the commutator action. For \(H\in\{H_d^0,H_d^-,H_u^+,H_u^0\}\), the right
action of the weak generators vanishes because
\begin{equation}
f_{+++-+}P=0,\qquad f_{+++--}P=0,
\end{equation}
while the left action survives since
\begin{equation}
PH=H.
\end{equation}
Therefore the commutator action reduces to a left action,
\begin{equation}
[T_i,H]=T_iH.
\end{equation}

Likewise we can determine the commutator action of $Q'$ and $Y$ on the Higgs components. The eigenvalues are summarised in Table \ref{tab:first_generation_Higgs_quantum_numbers}.
\begin{table}[htbp]
\centering
\renewcommand{\arraystretch}{1.2}
\begin{tabular}{c c c c c}
\toprule
Component & Algebraic representative & \(T_3\) & \(Q'\) & \(Y\) \\
\midrule
\(H_d^0\) & \(a_4 a_5^\dagger f_{+++-+}\) & \(+\tfrac12\) & \(0\) & \(-1\) \\
\(H_d^-\) & \(a_4 f_{+++-+}\) & \(-\tfrac12\) & \(-1\) & \(-1\) \\
\(H_u^+\) & \(a_4 f_{+++--}\) & \(+\tfrac12\) & \(+1\) & \(+1\) \\
\(H_u^0\) & \(a_4 a_5 f_{+++--}\) & \(-\tfrac12\) & \(0\) & \(+1\) \\
\bottomrule
\end{tabular}
\caption{Quantum numbers of the first-generation Higgs components.}
\label{tab:first_generation_Higgs_quantum_numbers}
\end{table}

The commutator relations above, together with Table~\ref{tab:first_generation_Higgs_quantum_numbers},
show that the two first-generation Higgs doublets transform under
\(SU(3)_C\times SU(2)_L\times U(1)_Y\) as
\begin{equation}
H_d=\begin{pmatrix}
H_d^0\\
H_d^-
\end{pmatrix}
\sim (1,2,-1),
\qquad
H_u=\begin{pmatrix}
H_u^+\\
H_u^0
\end{pmatrix}
\sim (1,2,+1).
\end{equation}

This is the same opposite-hypercharge \(H_u/H_d\) basis familiar from supersymmetric model building, where one works directly with two Higgs doublets carrying hypercharges \(+1\) and \(-1\), rather than starting from a single doublet and its conjugate.

The two doublets are algebraically distinct because \(H_d\subset I_9\) and \(H_u\subset I_{25}\) take values in different minimal-ideal sectors. They are not related by the usual electroweak conjugation \(i\sigma_2 H^*\), because neither complex conjugation nor left multiplication changes the primitive idempotents (and thus the minimal left ideal). At the level of the first-generation algebraic construction, the model yields a Type-II-like two-Higgs-doublet structure.

Although the two doublets are related by the purely internal algebraic operation of right multiplication,
\begin{eqnarray}
H_u = H_d a_5,\quad H_d = H_u a_5^{\dagger}
\end{eqnarray}
this does not identify them as the same Higgs field. Rather, it is merely an internal relation between two distinct Higgs doublets.

\section{Cyclic Triplication of the Higgs Sector}
\label{sec:triplication}

We now turn to one of the central structural consequences of the present construction: the first-generation Higgs doublets are not fixed under the order-three family generator $\psi_3$. Instead, they are mapped into two further pairs of Higgs doublets. The result is a family-resolved Higgs sector organised into cyclic $\psi_3$-orbits.

Let us first rewrite the first generation Higgs doublets as
\begin{equation}
H_d^{(1)}:=
\begin{pmatrix}
a_4 a_5^\dagger\,f_{+++-+} \\
a_4\,f_{+++-+}
\end{pmatrix},\qquad 
H_u^{(1)}:=
\begin{pmatrix}
a_4\,f_{+++--}\\
a_4 a_5\,f_{+++--}
\end{pmatrix}.
\end{equation}
Since \(\psi_3\) is an algebra automorphism, it acts on these doublets component-wise:
\begin{equation}
H_d^{(2)}:=\psi_3\!\left(H_d^{(1)}\right)=
\begin{pmatrix}
\psi_3(a_4 a_5^\dagger)\,\psi_3(f_{+++-+}) \\
\psi_3(a_4)\,\psi_3(f_{+++-+})
\end{pmatrix},
\end{equation}
\begin{equation}
H_u^{(2)}:=\psi_3\!\left(H_u^{(1)}\right)=
\begin{pmatrix}
\psi_3(a_4)\,\psi_3(f_{+++--}) \\
\psi_3(a_4 a_5)\,\psi_3(f_{+++--})
\end{pmatrix}.
\end{equation}
Likewise,
\begin{equation}
H_d^{(3)}:=\psi_3^2\!\left(H_d^{(1)}\right)=
\begin{pmatrix}
\psi_3^2(a_4 a_5^\dagger)\,\psi_3^2(f_{+++-+}) \\
\psi_3^2(a_4)\,\psi_3^2(f_{+++-+})
\end{pmatrix},
\end{equation}
\begin{equation}
H_u^{(3)}:=\psi_3^2\!\left(H_u^{(1)}\right)=
\begin{pmatrix}
\psi_3^2(a_4)\,\psi_3^2(f_{+++--}) \\
\psi_3^2(a_4 a_5)\,\psi_3^2(f_{+++--})
\end{pmatrix}.
\end{equation}

The family-resolved Higgs sector therefore consists of six doublets
\begin{equation}
\{H_d^{(1)},H_d^{(2)},H_d^{(3)}\},\qquad\{H_u^{(1)},H_u^{(2)},H_u^{(3)}\},
\end{equation}
that are permuted cyclically by \(\psi_3\):
\begin{equation}
H_d^{(1)} \xrightarrow{\ \psi_3\ } H_d^{(2)}\xrightarrow{\ \psi_3\ } H_d^{(3)} \xrightarrow{\ \psi_3\ } H_d^{(1)},
\end{equation}
\begin{equation}
H_u^{(1)} \xrightarrow{\ \psi_3\ } H_u^{(2)}\xrightarrow{\ \psi_3\ } H_u^{(3)} \xrightarrow{\ \psi_3\ } H_u^{(1)}.
\end{equation}
The \(\psi_3\)-images are distinct algebraic objects.

All three copies of each Higgs type carry the same gauge quantum numbers. All the gauge generators of the $\bb{C}\ell(10)$ model are \(S_3\)-invariant. Thus for any such gauge generator \(G\) and any algebra element \(X\),
\begin{equation}
\psi_3([G,X])=[\psi_3(G),\psi_3(X)]=[G,\psi_3(X)],
\end{equation}
and similarly for \(\psi_3^2\). It follows that if \(X\) is an eigenvector of the adjoint action of \(G\) with eigenvalue \(\lambda\), then \(\psi_3(X)\) is also an eigenvector with the same eigenvalue. Therefore all gauge quantum numbers are preserved along the \(\psi_3\)-orbit.

\subsection{Triplicated Higgs doublets}
It is instructive to display one transformed component explicitly. In the \(32\times32\) matrix representation, the first-generation down-type neutral component is
\begin{equation}
H_d^{0(1)} =a_4 a_5^\dagger f_{+++-+}= -\,E_{17,9}.
\end{equation}
Applying \(\psi_3\) gives
\begin{eqnarray}
\nonumber H_d^{0(2)}&:=&\psi_3\left(H_d^{0(1)}\right),\\
\nonumber &=&\left(\frac{1}{4}\,a_4 a_5^\dagger-\frac{\sqrt{3}}{4}\,a_1^\dagger a_2^\dagger a_3^\dagger a_4 a_5^\dagger\right) f_{+++-+}
+\left(-\frac{\sqrt{3}}{4}\,a_1 a_2 a_3 a_4 a_5^\dagger-\frac{3}{4}\,a_4 a_5^\dagger\right) f_{----+},\\
&=&-\frac14 E_{17,9}-\frac{i\sqrt{3}}{4} E_{30,9}-\frac{i\sqrt{3}}{4} E_{17,6}+\frac34 E_{30,6},
\end{eqnarray}
while
\begin{eqnarray}
\nonumber H_d^{0(3)}&:=&\psi_3^2\left(H_d^{0(1)}\right)\\
\nonumber &=&\left(\frac{1}{4}\,a_4 a_5^\dagger+\frac{\sqrt{3}}{4}\,a_1^\dagger a_2^\dagger a_3^\dagger a_4a_5^\dagger\right) f_{+++-+}
+\left(\frac{\sqrt{3}}{4}\,a_1 a_2 a_3 a_4 a_5^\dagger-\frac{3}{4}\,a_4a_5^\dagger\right) f_{----+},\\
&=&-\frac14 E_{17,9}+\frac{i\sqrt{3}}{4} E_{30,9}+\frac{i\sqrt{3}}{4} E_{17,6}
+\frac34 E_{30,6}.
\end{eqnarray}
We see that $H_d^{0(2)}$ and $H_d^{0(3)}$ have support in two columns (minimal left ideals), and that
\begin{equation}
\operatorname{span}_{\bb{C}}\!\left\{H_d^{0(1)},\,H_d^{0(2)},\,H_d^{0(3)}\right\}\subset I_{6}\oplus I_{9},
\end{equation}
where $I_{6}=\bb{C}\ell(10)f_{----+}$ and $I_{9}=\bb{C}\ell(10)f_{+++-+}$ are the minimal left ideals corresponding to the 6th and 9th column in the $\text{Mat}(32,\bb{C})$ representation of $\bb{C}\ell(10)$. 

An analogous computation for the up-type Higgs gives components supported in \(I_{22}\oplus I_{25}\), so that
\begin{equation}
\operatorname{span}_{\bb{C}}\left\{H_u^{+(1)},\,H_u^{+(2)},\,H_u^{+(3)}\right\}\subset I_{22}\oplus I_{25}.
\end{equation}
A full list of the three $\psi_3$-related Higgs components in the Witt basis is given in Appendix~\ref{app:Higgs}.

\subsection{Linear independence of the triplicated Higgs sector}
\label{sec:independence}

The fermion sector is \(96\)-dimensional \cite{gourlay2024algebraic}, with
\begin{equation}
\dim\!\left(S_1\oplus S_2\oplus S_3\right)=96,
\end{equation}
where
\begin{equation}
S_1=V_1^+\oplus V_1^+a_5\oplus V_1^+a_4\oplus V_1^+a_4a_5,\qquad
S_2=\psi_3(S_1),\qquad
S_3=\psi_3^2(S_1).
\end{equation}

The family-resolved Higgs sector contributes twelve further linearly independent states,
\begin{equation}
\dim\;\operatorname{span}_{\bb{C}}\!\left(\{H_d^{0(r)},H_d^{-(r)},H_u^{+(r)},H_u^{0(r)}\}_{r=1,2,3}\right)=12.
\end{equation}
Furthermore, one verifies in Mathematica \cite{VerificationRepository2} that these twelve Higgs states are linearly independent from the \(96\) fermion states. Hence the combined fermion-Higgs sector has dimension \(108\).

\subsection{Symmetric and orthogonal Higgs combinations}

For each Higgs component, the three family copies
\begin{eqnarray}
H^{(1)},\qquad H^{(2)}=\psi_3(H^{(1)}),\qquad H^{(3)}=\psi_3^2(H^{(1)}),
\end{eqnarray}
are cyclically permuted by the action of \(\psi_3\). It is therefore natural to form their orbit average, in direct analogy with the construction of the \(S_3\)-invariant electric charge generator \(Q'\) in \eqref{eq:Q'def} from the first-generation operator \eqref{eq:Q1def}. We therefore define
\begin{eqnarray}
    \bar{H}_d^0&:=&\frac{1}{3}\left(H_d^{0(1)}+H_d^{0(2)}+H_d^{0(3)}\right),\\
    \bar{H}_d^-&:=&\frac{1}{3}\left(H_d^{-(1)}+H_d^{-(2)}+H_d^{-(3)}\right),\\
    \bar{H}_u^+&:=&\frac{1}{3}\left(H_u^{+(1)}+H_u^{+(2)}+H_u^{+(3)}\right),\\
    \bar{H}_u^0&:=&\frac{1}{3}\left(H_u^{0(1)}+H_u^{0(2)}+H_u^{0(3)}\right).
\end{eqnarray}

Algebraically, the resulting symmetric Higgs doublets are
\begin{eqnarray}
    \bar{H}_d=\begin{pmatrix}
\bar{H}_d^0 \\
\bar{H}_d^-
\end{pmatrix}
=
\begin{pmatrix}
\frac{1}{2}a_4a_5^{\dagger}(f_{+++-+}-f_{----+}) \\
\frac{1}{2}a_4(f_{+++-+}+f_{----+})
\end{pmatrix},
\end{eqnarray}
and
\begin{eqnarray}
    \bar{H}_u=\begin{pmatrix}
\bar{H}_u^+ \\
\bar{H}_u^0
\end{pmatrix}
=
\begin{pmatrix}
\frac{1}{2}a_4(f_{+++--}+f_{-----}) \\
\frac{1}{2}a_4a_5(f_{+++--}-f_{-----})
\end{pmatrix}.
\end{eqnarray}
The notation $\bar H$ denotes the average over the order-three $\psi_3$ orbit. These averaged Higgs components are invariant under $\psi_3$. Under the order-two generator $\epsilon$, however, they acquire an overall minus sign: $\epsilon(\bar H)=-\bar H$. Thus the order-three orbit average is not a singlet under the full $S_3$ action, but transforms in the sign representation $1'$.

Rather than working with the full family-resolved Higgs sector, we now restrict attention to these cyclic combinations. The remaining orthogonal orbit directions are still present algebraically, but will not be needed in the cyclically averaged Higgs limit considered here. They become important once \(S_3\) breaking and flavour phenomenology are introduced, and a detailed analysis of that sector lies beyond the scope of the present paper.

The three \(\psi_3\)-related Higgs copies may also be decomposed into two orthogonal directions. For each Higgs component \(H\in\{H_d^0,H_d^-,H_u^+,H_u^0\}\), define
\begin{eqnarray}
H_\alpha &:=& H^{(1)} - \frac{1}{2}\bigl(H^{(2)} + H^{(3)}\bigr),\\
H_\beta &:=& \frac{\sqrt{3}}{2}\bigl(H^{(2)} - H^{(3)}\bigr).
\end{eqnarray}
Under \(\psi_3\), the pair \((H_\alpha,H_\beta)\) transforms as a rotation by \(2\pi/3\). In the present paper we restrict attention to the cyclically averaged Higgs pair \(\bar H_d,\bar H_u\). The orthogonal orbit directions will become relevant once \(S_3\) breaking and realistic flavour structure are studied.

\subsection{Effect of passing to the cyclically averaged Higgs limit}

When the fermions are kept in the first-generation subspaces, replacing the family-resolved Higgs operators by the cyclically averaged Higgs doublets does not change the set of allowed Yukawa channels. The reason is not only the linearity of
\begin{eqnarray}
\mathcal{Y}(X_L,H,X_R)=\Tr\bigl((X_LH)^\dagger X_R\bigr),
\end{eqnarray}
but also the ideal support of the \(\psi_3\)-images.

For example, consider the first-generation charged lepton channel
\begin{equation}
e_L^- = a_4^\dagger f_{++++-}\in V_1^-,
\qquad
e_R^- = a_5^\dagger f_{+++-+}\in U_1^-.
\end{equation}
For the first-generation Higgs component one finds
\begin{eqnarray}
e_L^- H_d^{0(1)}&=&a_5^\dagger f_{+++-+}=e_R^-.
\end{eqnarray}
For the second Higgs copy one obtains
\begin{eqnarray}
\nonumber e_L^- H_d^{0(2)}
&=&
a_4^\dagger f_{++++-}
\Bigl[
\Bigl(\frac14 a_4a_5^\dagger-\frac{\sqrt3}{4}a_1^\dagger a_2^\dagger a_3^\dagger a_4a_5^\dagger\Bigr)f_{+++-+}\\
\nonumber &&\hspace{1.2cm}+
\Bigl(-\frac{\sqrt3}{4}a_1a_2a_3a_4a_5^\dagger-\frac34 a_4a_5^\dagger\Bigr)f_{----+}
\Bigr] \\
\nonumber &=&
\frac14\,a_5^\dagger f_{+++-+}
+\frac{\sqrt3}{4}\,a_1a_2a_3a_5^\dagger f_{----+},\\
&=& \frac14\,e_R^-+\frac{\sqrt3}{4}\,a_1a_2a_3a_5^\dagger f_{----+}.
\end{eqnarray}
Similarly,
\begin{eqnarray}
e_L^- H_d^{0(3)}=\frac14\,e_R^-
-\frac{\sqrt3}{4}\,a_1a_2a_3a_5^\dagger f_{----+}.
\end{eqnarray}
Hence
\begin{eqnarray}
e_L^- \bar H_d^0
&=&
\frac13\Bigl(e_L^- H_d^{0(1)}+e_L^- H_d^{0(2)}+e_L^- H_d^{0(3)}\Bigr)= \frac12\,e_R^-,
\end{eqnarray}
and therefore
\begin{eqnarray}
\mathcal{Y}(e_L^-,\bar H_d^0,e_R^-)=\frac12.
\end{eqnarray}

Thus passing from \(H_d^{0(1)}\) to the cyclically averaged component \(\bar H_d^0\) rescales the Yukawa coefficient from \(1\) to \(\frac12\), but does not change the Yukawa channel itself. The same reasoning applies to the other Higgs components. In particular, \(\bar H_d\) still couples only to the \(U_1^-\)-type singlet subspace, while \(\bar H_u\) still couples only to the \(U_1^+\)-type singlet subspace. Hence replacing the family-resolved Higgs sector by the cyclically averaged Higgs doublets does not introduce any new wrong-type Yukawa couplings and does not change the allowed-versus-forbidden pattern of Yukawa interactions.

For these first-generation channels, what changes is the numerical value of the Yukawa coefficient: the cyclically averaged Higgs component is an average over the \(\psi_3\)-orbit, so the corresponding coefficient is modified by the surviving projection onto the relevant source and target ideals. At the present stage this is not yet of physical importance, since our goal is only to determine which Yukawa channels are allowed and which are forbidden, rather than to extract fermion masses or mixing angles.

\section{Flavour Structure in the Cyclically Averaged Higgs Limit}
\label{sec:flavour_discussion}

\subsection{Type-II structure in the cyclically averaged Higgs limit}

In generic multi-Higgs models, tree-level FCNCs are generally present unless the Yukawa sector has special structure. A standard sufficient criterion for natural flavour conservation is the Glashow--Weinberg condition: fermions of a given charge sector should couple to only one Higgs doublet. In the present construction, the relevant comparison is with a Type-II pattern.

At the level of the two first-generation Higgs doublets, the algebra clearly separates the Yukawa channels:
\begin{eqnarray}
H_d^{(1)} \text{ maps to the singlet subspace } U_1^- \subset I_9,\\
H_u^{(1)} \text{ maps to the singlet subspace } U_1^+ \subset I_{25}.
\end{eqnarray}
Thus down-type quarks and charged leptons couple through \(H_d^{(1)}\), whereas up-type quarks and neutrinos couple through \(H_u^{(1)}\). In this sense, the first-generation Higgs sector is naturally Type-II-like.

Once the Higgs sector is triplicated by \(\psi_3\), however, this simple Type-II-like picture is obscured: in the algebraic generation basis \((S_1,S_2,S_3)\), with \(S_2=\psi_3(S_1)\) and \(S_3=\psi_3^2(S_1)\), the Higgs triplication is not generation-diagonal, so the full family-resolved Higgs sector is not simply three decoupled copies of a two-Higgs-doublet model.

For the cyclically averaged Higgs doublets, by contrast, the relevant right-action maps satisfy
\begin{eqnarray}
R_{\bar H_d^-}\bigl(\psi_3^i(V_1^+)\bigr)\subseteq\psi_3^i(U_1^-),\qquad
R_{\bar H_d^0}\bigl(\psi_3^i(V_1^-)\bigr)\subseteq\psi_3^i(U_1^-),\\
R_{\bar H_u^0}\bigl(\psi_3^i(V_1^+)\bigr)\subseteq\psi_3^i(U_1^+),\qquad
R_{\bar H_u^+}\bigl(\psi_3^i(V_1^-)\bigr)\subseteq\psi_3^i(U_1^+),
\end{eqnarray}
where \(i=0,1,2\). At the level of these right-action maps, the cyclically averaged Higgs doublets preserve the Type-II structure already identified for the first-generation Higgs sector: \(\bar H_d\) couples only to the down-type singlet subspace, whereas \(\bar H_u\) couples only to the up-type singlet subspace. Thus, in the cyclically averaged Higgs limit, the algebraic Yukawa structure is again Type-II-like, even though, as we now discuss, the resulting generation-space Yukawa matrices are not diagonal in the algebraic generation basis.

\subsection{Generation-space Yukawa matrices}

The Yukawa matrices are not diagonal in the algebraic generation basis. This is because, although the states in the three subspaces \(S_1\), \(S_2\), and \(S_3\) are linearly independent, they are not mutually orthogonal under the Hilbert--Schmidt pairing.

This can be seen explicitly in the down-type quark sector. One finds
\begin{eqnarray}
\left\langle d_{R,1}^{(i)},\,d_{R,2}^{(j)}\right\rangle&:=&\Tr\left(\left(d_{R,1}^{(i)}\right)^\dagger d_{R,2}^{(j)}\right)=\frac14\,\delta_{ij}.
\end{eqnarray}
On the other hand, for the cyclically averaged down-type Higgs one has
\begin{eqnarray}
u_{L,1}^{(i)}\,\bar H_d^-=\frac12\,d_{R,1}^{(i)}.
\end{eqnarray}
Therefore
\begin{eqnarray}
\nonumber\mathcal{Y}\!\left(u_{L,1}^{(i)},\,\bar H_d^{-},\,d_{R,2}^{(j)}\right)
&=&
\Tr\left(\left(u_{L,1}^{(i)}\bar H_d^-\right)^\dagger d_{R,2}^{(j)}\right)\\
&=&
\nonumber\Tr\left(\left(\frac12\,d_{R,1}^{(i)}\right)^\dagger d_{R,2}^{(j)}\right)\\
&=&
\frac18\,\delta_{ij}.
\end{eqnarray}
Thus the cyclically averaged Higgs does not mix generations at the level of the right-action map, but the resulting state can still have nonzero overlap with fermion states in the other \(\psi_3\)-related subspaces.

For definiteness, consider again the down-type quark sector. Define the generation-space Yukawa matrices by
\begin{eqnarray}
\mathcal{Y}\!\left(u_{L,r}^{(i)},\,H_d^{-(a)},\,d_{R,s}^{(j)}\right)
=(M_a)_{rs}\,\delta_{ij},
\qquad a,r,s=1,2,3.
\end{eqnarray}
Evaluating the pairings gives
\begin{eqnarray}
M_1&=&
\begin{pmatrix}
1 & \frac14 & \frac14 \\
\frac14 & \frac14 & -\frac18 \\
\frac14 & -\frac18 & \frac14
\end{pmatrix},\quad
M_2=
\begin{pmatrix}
\frac14 & \frac14 & -\frac18 \\
\frac14 & 1 & \frac14 \\
-\frac18 & \frac14 & \frac14
\end{pmatrix},\quad
M_3=
\begin{pmatrix}
\frac14 & -\frac18 & \frac14 \\
-\frac18 & \frac14 & \frac14 \\
\frac14 & \frac14 & 1
\end{pmatrix}.
\end{eqnarray}
Linearity of the Yukawa pairing then gives
\begin{eqnarray}
(\bar M)_{rs}:=\mathcal{Y}\!\left(u_{L,r}^{(i)},\,\bar H_d^{-},\,d_{R,s}^{(j)}\right)
=\frac13\Bigl((M_1)_{rs}+(M_2)_{rs}+(M_3)_{rs}\Bigr),
\end{eqnarray}
and hence
\begin{eqnarray}
\bar M=\frac13(M_1+M_2+M_3)=
\begin{pmatrix}
\frac12 & \frac18 & \frac18 \\
\frac18 & \frac12 & \frac18 \\
\frac18 & \frac18 & \frac12
\end{pmatrix}.
\end{eqnarray}

Thus, in the algebraic generation basis \((S_1,S_2,S_3)\), the averaged Yukawa matrix is not diagonal. The explicit nonzero entry
\begin{eqnarray}
\mathcal{Y}\!\left(u_{L,1}^{(i)},\,\bar H_d^{-},\,d_{R,2}^{(j)}\right)=\frac18\,\delta_{ij}
\end{eqnarray}
is precisely one example of this off-diagonal structure.

The same \(S_3\)-invariant texture \(\bar M\) arises in the other allowed Yukawa sectors as well, up to overall sign conventions determined by the algebraic representatives. Since the algebraic construction fixes the matrix texture but not the overall sector-dependent Yukawa constants, these signs do not affect the flavour structure. In the cyclically averaged Higgs limit, it is therefore natural to parameterise the allowed Yukawa sectors by separate overall couplings \(y_d\) and \(y_u\) for the down-type and up-type quark sectors, together with the corresponding Higgs vacuum expectation values \(v_d\) and \(v_u\). Analogous overall constants appear in the charged-lepton and neutrino sectors. Thus the nontrivial result of the present construction is the algebraic determination of the matrix texture itself, rather than the absolute mass scales.

\subsection{Eigenvalues and comparison with conventional $S_3$ textures}

The averaged Yukawa matrix \(\bar{M}\) can be written as
\begin{equation}
\bar{M}=\tfrac{3}{8}\mathbf{I}_3+\tfrac{1}{8}\mathbf{J}_3,
\end{equation}
where \(\mathbf{I}_3\) is the \(3\times 3\) identity matrix and \(\mathbf{J}_3\) is the \(3\times 3\) all-ones matrix. 

This is readily diagonalised. The all-ones matrix \(\mathbf{J}_3\) has eigenvalue $3$ on the family-symmetric line spanned by $(1,1,1)$ and eigenvalue $0$ on the orthogonal two-dimensional subspace. Hence $\bar M$ has eigenvalues \begin{eqnarray} \lambda_{\rm sym}=\frac{3}{8}+\frac{1}{8}\cdot 3=\frac{3}{4}, \qquad \lambda_{\perp}=\frac{3}{8}, \end{eqnarray} where $\lambda_{\perp}$ is doubly degenerate. Choosing the unit-normalised representative, the eigenvector corresponding to $\lambda_{\rm sym}=3/4$ is $(1,1,1)/\sqrt{3}$. Thus, if electroweak symmetry breaking is implemented in the usual way so that the corresponding mass matrix is proportional to $\bar M$, the cyclically averaged Higgs limit yields a mass pattern with ratios $2:1:1$ up to an overall sector-dependent scale. In this sense, the cyclically averaged Higgs limit distinguishes one family-symmetric combination and leaves a degenerate pair. This indicates that additional $S_3$-breaking effects, or vacuum expectation values in the orthogonal Higgs directions, are needed to lift the degeneracy.

It is instructive to compare \(\bar{M}\) with the democratic mass matrix
\(M_{\mathrm{dem}}=c\,\mathbf{J}_3\) familiar from the \(S_3\) flavour literature
\cite{mondragon2007lepton,babu2024fermion}. The democratic matrix has eigenvalues \((3c,0,0)\), giving two exactly massless generations in the symmetric limit. By contrast, \(\bar{M}=\tfrac{3}{8}\mathbf{I}_3+\tfrac{1}{8}\mathbf{J}_3\) has eigenvalues \((\tfrac{3}{4},\tfrac{3}{8},\tfrac{3}{8})\), so no generation is massless in the cyclically averaged Higgs limit. The crucial difference is the identity component \(\tfrac{3}{8}\mathbf{I}_3\), which arises from the intra-generation trace pairings. Moreover, the ratio of diagonal to off-diagonal entries is fixed at \(4:1\) by the algebraic construction. In many conventional \(S_3\) flavour models, by contrast, such a ratio is not fixed by the family symmetry alone, but depends on the Yukawa parameters, Higgs assignments, and vacuum expectation values \cite{mondragon2007lepton,gonzalez2013quark,teshima2012higgs,cogollo2016two}.

It is also worth noting that \(\bar M\) is not of the fully democratic form, for which all entries are equal and the eigenvalues are proportional to \((3,0,0)\). Instead, the diagonal and off-diagonal entries take the specific values \(\frac12\) and \(\frac18\), fixed by the algebraic construction. In this sense the present framework differs from conventional \(S_3\) flavour models, in which the texture is typically parametrised phenomenologically and the matrix entries remain free up to symmetry constraints. Here the relative pattern of entries is not chosen by hand but follows directly from the underlying algebra.

\subsection{FCNCs in the cyclically averaged Higgs limit}

The nonzero off-diagonal entries in the algebraic generation basis do not by themselves imply tree-level FCNCs. In the cyclically averaged Higgs limit there is only one neutral Yukawa matrix for each fermion charge sector. If electroweak symmetry breaking is implemented in the usual way, so that the neutral component of the corresponding Higgs doublet acquires a vacuum expectation value, then the fermion mass matrix for that sector is proportional to this same Yukawa matrix. The neutral Higgs couplings are therefore aligned with the mass matrix, and are diagonalised by the same change of fermion basis. Thus, in the cyclically averaged Higgs limit, no tree-level FCNCs are expected to arise. At the same time, this cyclically averaged Higgs limit is likely too restrictive to account for realistic flavour, thereby motivating the need for \(S_3\)-breaking.

A more complete understanding of this limit will require an analysis of the scalar potential and vacuum structure in the cyclically averaged and orthogonal Higgs directions. We leave this to future work. 

\section{Discussion and Outlook}

The present paper constructs the Higgs and Yukawa sectors of an algebraic three-generation model based on $\bb{C}\ell(10)$ with $S_3$ family symmetry. The motivation was to address a common limitation of algebraic SM constructions: while such frameworks often give natural descriptions of SM fermion multiplets and gauge generators, the Higgs and Yukawa sectors remain underdeveloped or absent. Here we have shown that, in the present model, Higgs degrees of freedom can be realised internally as right-action operators mapping weak-doublet fermion subspaces into the corresponding weak-singlet subspaces.

Starting from the first-generation fermion subspaces, we identified two Higgs doublets with electroweak quantum numbers $(1,2,-1)$ and $(1,2,+1)$. The corresponding Yukawa coefficients were extracted using a Hilbert--Schmidt trace pairing. This construction gives a Type-II-like separation of Yukawa channels: the down-type Higgs doublet maps to the down-type singlet subspace, while the up-type Higgs doublet maps to the up-type singlet subspace. Acting with the order-three family generator $\psi_3$ then generates a family-resolved Higgs sector organised into $S_3$ orbits. The resulting Higgs states are linearly independent, and their cyclic averages define the cyclically averaged Higgs doublets studied in this paper.

In this cyclically averaged Higgs limit, the Type-II-like Yukawa selection rule is preserved. At the same time, the generation-space Yukawa matrices are not diagonal in the algebraic generation basis, because the three algebraic generation subspaces are linearly independent but not mutually orthogonal under the trace pairing. The resulting cyclically averaged Yukawa texture is fixed algebraically, with diagonal entries $\frac12$ and off-diagonal entries $\frac18$. Equivalently, the matrix contains a family-symmetric eigendirection and a degenerate orthogonal pair. Thus the construction does not merely reproduce a generic multi-Higgs setup: it determines a specific flavour texture from the underlying algebraic structure.

This fixed texture should be understood as a structural result rather than as a complete realistic flavour model. If electroweak symmetry breaking is implemented in the usual way, so that the fermion mass matrix in each charge sector is proportional to the corresponding neutral Yukawa matrix, then the neutral Higgs couplings are aligned with the mass matrices. Consequently, tree-level FCNCs are not expected in the cyclically averaged Higgs limit. However, the cyclically averaged Higgs limit leaves a degenerate pair of fermion directions, and therefore cannot by itself account for the observed pattern of fermion masses, mixing angles, and CP violation. Realistic flavour requires going beyond this limit.

The natural next step is to study how $S_3$ breaking is implemented in the Higgs sector. The present construction already contains, before orbit averaging, a family-resolved Higgs structure with cyclically averaged and orthogonal orbit directions. This makes it possible to ask whether vacuum expectation values in the orthogonal Higgs directions, soft $S_3$ breaking, or a nontrivial scalar potential can generate realistic Yukawa textures in a way that is more constrained than in conventional phenomenological flavour models. This connects the present framework directly with the existing $S_3$ flavour literature, where realistic masses and mixings typically require nontrivial Higgs-family structure, vacuum alignment, or explicit or soft breaking of the family symmetry \cite{teshima2012higgs,canales2013fermion,gonzalez2013s3,kaneko2007flavor}.

A further direction is to compare the present algebraic Higgs construction with tri-hypercharge models \cite{navarro2023tri,fernandez2024minimal}. In those models, family-dependent weak hypercharges and staged symmetry breaking provide a route toward flavour structure. Since the present algebraic framework also contains family-resolved gauge and Higgs structures before passing to averaged or generation-independent combinations, it is natural to ask whether analogous flavour-generating mechanisms can emerge here from the algebra itself. The results of this paper therefore provide a structural Higgs/Yukawa completion of the $\bb{C}\ell(10)$ three-generation framework and establish a concrete starting point for future work on $S_3$ breaking, scalar potentials, and realistic flavour phenomenology.

\subsubsection*{Acknowledgements}
The author is particularly grateful to Janek Kozicki for his independent Mathematica verification of the calculations presented in this work \cite{VerificationRepository2}.

\section*{Declarations}

\subsection*{Funding}
No funding was received to assist with the preparation of this manuscript.

\subsection*{Competing interests}
The author has no relevant financial or non-financial interests to disclose.

\subsection*{Data and code availability}
No experimental datasets were generated or analysed in this study. The Mathematica verification notebook supporting the algebraic calculations is publicly available in the GitLab repository cited in Ref.~\cite{VerificationRepository2}.

\bibliographystyle{unsrt}
\bibliography{bibliography.bib}
\newpage

\appendix
\section{Algebraic Identification of First Generation of Fermions}\label{sec:appA}

The first-generation algebraic state assignments used throughout the main text are summarised in Table~\ref{tab:first-generation-states}. The states are written in terms of the Witt basis of $\bb{C}\ell(10)$ and are identified by their eigenvalues under $T_3$, $Q'$, and $Y$, computed using the commutator action of the gauge generators. The hypercharge is defined by $Y=2(Q'-T_3)$. Throughout the table, $i,j\in\{1,2,3\}$ with $i<j$, and superscripts $(3)$ and $(\bar 3)$ denote colour triplet and antitriplet representations of $SU(3)_C$.

\begin{table}[h!]
\centering
\renewcommand{\arraystretch}{1.5}
\begin{tabular}{|c|c|c|c|c|}
\hline
State & $T_3$ eigenvalue & $Q'$ eigenvalue & $Y$ eigenvalue & Particle \\ \hline

$a_1^{\dagger}a_2^{\dagger}a_3^{\dagger}a_4^{\dagger}f_{+++++}$ 
& $+\frac{1}{2}$ 
& $1$ 
& $1$ 
& $e^+_L$ \\ \hline

$a_i^{\dagger}a_j^{\dagger}f_{+++++}$ 
& $+\frac{1}{2}$ 
& $+\frac{1}{3}$ 
& $-\frac{1}{3}$ 
& $\bar{d}_L^{(\bar{3})}$ \\ \hline

$a_i^{\dagger}a_4^{\dagger}f_{+++++}$ 
& $+\frac{1}{2}$  
& $+\frac{2}{3}$ 
& $+\frac{1}{3}$ 
& $u_L^{(3)}$ \\ \hline

$a_4^{\dagger}a_5^{\dagger}f_{+++++}$ 
& $+\frac{1}{2}$  
& $0$ 
& $-1$ 
& $\nu_L$ \\ 

\hline\hline

$a_1^{\dagger}a_2^{\dagger}a_3^{\dagger}a_4^{\dagger}a_5f_{++++-}$ 
& $-\frac{1}{2}$ 
& $0$ 
& $1$ 
& $\bar{\nu}_L$ \\ \hline

$a_i^{\dagger}a_j^{\dagger}a_5f_{++++-}$ 
& $-\frac{1}{2}$ 
& $-\frac{2}{3}$ 
& $-\frac{1}{3}$ 
& $\bar{u}_L^{(\bar{3})}$ \\ \hline

$a_i^{\dagger}a_4^{\dagger}a_5f_{++++-}$ 
& $-\frac{1}{2}$  
& $-\frac{1}{3}$ 
& $+\frac{1}{3}$ 
& $d_L^{(3)}$ \\ \hline

$a_4^{\dagger}f_{++++-}$ 
& $-\frac{1}{2}$  
& $-1$ 
& $-1$ 
& $e^-_L$ \\ 

\hline\hline

$a_1^{\dagger}a_2^{\dagger}a_3^{\dagger}a_5f_{+++--}$ 
& $0$ 
& $1$ 
& $+2$ 
& $e^+_R$ \\ \hline

$a_i^{\dagger}a_j^{\dagger}a_4a_5f_{+++--}$ 
& $0$ 
& $+\frac{1}{3}$ 
& $+\frac{2}{3}$ 
& $\bar{d}_R^{(\bar{3})}$ \\ \hline

$a_i^{\dagger}a_5f_{+++--}$ 
& $0$  
& $+\frac{2}{3}$ 
& $+\frac{4}{3}$ 
& $u_R^{(3)}$ \\ \hline

$f_{+++--}$ 
& $0$  
& $0$ 
& $0$ 
& $\nu_R$ \\ 

\hline\hline

$a_1^{\dagger}a_2^{\dagger}a_3^{\dagger}f_{+++-+}$ 
& $0$ 
& $0$ 
& $0$ 
& $\bar{\nu}_R$ \\ \hline

$a_i^{\dagger}a_j^{\dagger}a_4f_{+++-+}$ 
& $0$ 
& $-\frac{2}{3}$ 
& $-\frac{4}{3}$ 
& $\bar{u}_R^{(\bar{3})}$ \\ \hline

$a_i^{\dagger}f_{+++-+}$ 
& $0$  
& $-\frac{1}{3}$ 
& $-\frac{2}{3}$ 
& $d_R^{(3)}$ \\ \hline

$a_5^{\dagger}f_{+++-+}$ 
& $0$  
& $-1$ 
& $-2$ 
& $e^-_R$ \\ \hline

\end{tabular}
\caption{Eigenvalues of $T_3$, $Q'$, and $Y$ for the first generation of algebraic states, identified with the first generation of fermions.}
\label{tab:first-generation-states}
\end{table}

\newpage
\section{Summary Table of Allowed and Forbidden Yukawa Channels}
For the first-generation Higgs components, the allowed and forbidden Yukawa channels are summarised in Tables~\ref{tab:allowed_yukawa_channels} and \ref{tab:forbidden_yukawa_channels}. The same pattern is inherited by the \(\psi_3\)-images and by the cyclically averaged Higgs doublets, with only the nonzero coefficients being rescaled in the latter case.

\begin{table}[h]
\centering
\renewcommand{\arraystretch}{1.2}
\begin{tabular}{|c|c|c|c|}
\hline
\(X_L\) & \(H\) & \(X_R\) & Yukawa pairing \\
\hline
\(u_{L}^{(i)}\) & \(H_d^{-(1)}\) & \(d_{R}^{(j)}\) & \(\neq 0\), \(\propto \delta_{ij}\) \\
\(d_{L}^{(i)}\) & \(H_d^{0(1)}\) & \(d_{R}^{(j)}\) & \(\neq 0\), \(\propto \delta_{ij}\) \\
\(\nu_L\) & \(H_d^{-(1)}\) & \(e_R\) & \(\neq 0\) \\
\(e_L\) & \(H_d^{0(1)}\) & \(e_R\) & \(\neq 0\) \\
\hline
\(u_{L}^{(i)}\) & \(H_u^{0(1)}\) & \(u_{R}^{(j)}\) & \(\neq 0\), \(\propto \delta_{ij}\) \\
\(d_{L}^{(i)}\) & \(H_u^{+(1)}\) & \(u_{R}^{(j)}\) & \(\neq 0\), \(\propto \delta_{ij}\) \\
\(\nu_L\) & \(H_u^{0(1)}\) & \(\nu_R\) & \(\neq 0\) \\
\(e_L\) & \(H_u^{+(1)}\) & \(\nu_R\) & \(\neq 0\) \\
\hline
\end{tabular}
\caption{Allowed first-generation Yukawa channels for the first-generation Higgs components.}
\label{tab:allowed_yukawa_channels}
\end{table}

\begin{table}[h]
\centering
\renewcommand{\arraystretch}{1.2}
\begin{tabular}{|c|c|c|c|}
\hline
\(X_L\) & \(H\) & \(X_R\) & Yukawa pairing \\
\hline
\(u_{L}^{(i)}\) & \(H_d^{-(1)}\) & \(u_{R}^{(j)}\) & \(=0\) \\
\(d_{L}^{(i)}\) & \(H_d^{0(1)}\) & \(u_{R}^{(j)}\) & \(=0\) \\
\(\nu_L\) & \(H_d^{-(1)}\) & \(\nu_R\) & \(=0\) \\
\(e_L\) & \(H_d^{0(1)}\) & \(\nu_R\) & \(=0\) \\
\hline
\(u_{L}^{(i)}\) & \(H_u^{0(1)}\) & \(d_{R}^{(j)}\) & \(=0\) \\
\(d_{L}^{(i)}\) & \(H_u^{+(1)}\) & \(d_{R}^{(j)}\) & \(=0\) \\
\(\nu_L\) & \(H_u^{0(1)}\) & \(e_R\) & \(=0\) \\
\(e_L\) & \(H_u^{+(1)}\) & \(e_R\) & \(=0\) \\
\hline
\end{tabular}
\caption{Forbidden first-generation wrong-type Yukawa channels for the first-generation Higgs components.}
\label{tab:forbidden_yukawa_channels}
\end{table}

\newpage

\section{$\psi_3$ Images of Higgs States}\label{app:Higgs}

The following equations list the $\psi_3$-orbit of the first-generation Higgs components used in the main text. The superscript $(k)$ denotes the $k$th algebraic generation sector, with $H^{(1)}=H$, $H^{(2)}=\psi_3(H)$, and $H^{(3)}=\psi_3^2(H)$. All expressions are written back in the original $a_i,a_i^\dagger$ Witt basis of $\bb{C}\ell(10)$, with the idempotents $f_{\pm\pm\pm\pm\pm}$ defined as in Section~\ref{sec:algebraicsetup}. These formulae make explicit the family-resolved Higgs states whose cyclic averages define $\bar H_d$ and $\bar H_u$.

\small
\noindent\makebox[\textwidth][l]{%
\hspace*{-1.2cm}%
\begin{minipage}{1.1\textwidth}
\begin{eqnarray}
H_d^{0(1)} &:=& a_4 a_5^\dagger f_{+++-+}, \\
H_d^{0(2)} &:=&\left(\frac{1}{4}\,a_4 a_5^\dagger-\frac{\sqrt{3}}{4}\,a_1^\dagger a_2^\dagger a_3^\dagger a_4 a_5^\dagger\right) f_{+++-+}+\left(-\frac{\sqrt{3}}{4}\,a_1 a_2 a_3 a_4 a_5^\dagger-\frac{3}{4}\,a_4 a_5^\dagger\right) f_{----+}, \\
H_d^{0(3)} &:=&\left(\frac{1}{4}\,a_4 a_5^\dagger+\frac{\sqrt{3}}{4}\,a_1^\dagger a_2^\dagger a_3^\dagger a_4a_5^\dagger\right) f_{+++-+}+\left(\frac{\sqrt{3}}{4}\,a_1 a_2 a_3 a_4 a_5^\dagger-\frac{3}{4}\,a_4a_5^\dagger\right) f_{----+}, \\
H_d^{-(1)} &:=& a_4 f_{+++-+}, \\
H_d^{-(2)} &:=&\left(\frac{1}{4}\,a_4-\frac{\sqrt{3}}{4}\,a_1^\dagger a_2^\dagger a_3^\dagger a_4\right) f_{+++-+}+
\left(\frac{\sqrt{3}}{4}\,a_1 a_2 a_3 a_4+\frac{3}{4}\,a_4\right) f_{----+}, \\
H_d^{-(3)} &:=&\left(\frac{1}{4}\,a_4+\frac{\sqrt{3}}{4}\,a_1^\dagger a_2^\dagger a_3^\dagger a_4\right) f_{+++-+}+
\left(-\frac{\sqrt{3}}{4}\,a_1 a_2 a_3 a_4+\frac{3}{4}\,a_4\right) f_{----+}, \\
H_u^{+(1)} &:=& a_4 f_{+++--}, \\
H_u^{+(2)} &:=&\left(\frac{1}{4}\,a_4-\frac{\sqrt{3}}{4}\,a_1^\dagger a_2^\dagger a_3^\dagger a_4\right) f_{+++--}+
\left(\frac{\sqrt{3}}{4}\,a_1 a_2 a_3 a_4+\frac{3}{4}\,a_4\right) f_{-----}, \\
H_u^{+(3)} &:=&\left(\frac{1}{4}\,a_4+\frac{\sqrt{3}}{4}\,a_1^\dagger a_2^\dagger a_3^\dagger a_4\right) f_{+++--}+
\left(-\frac{\sqrt{3}}{4}\,a_1 a_2 a_3 a_4+\frac{3}{4}\,a_4\right) f_{-----}, \\
H_u^{0(1)} &:=& a_4 a_5 f_{+++--}, \\
H_u^{0(2)} &:=&\left(\frac{1}{4}\,a_4 a_5-\frac{\sqrt{3}}{4}\,a_1^\dagger a_2^\dagger a_3^\dagger a_4 a_5\right)f_{+++--}+\left(-\frac{\sqrt{3}}{4}\,a_1 a_2 a_3 a_4 a_5-\frac{3}{4}\,a_4 a_5\right) f_{-----}, \\
H_u^{0(3)} &:=&\left(\frac{1}{4}\,a_4 a_5+\frac{\sqrt{3}}{4}\,a_1^\dagger a_2^\dagger a_3^\dagger a_4 a_5\right)f_{+++--}+\left(\frac{\sqrt{3}}{4}\,a_1 a_2 a_3 a_4 a_5-\frac{3}{4}\,a_4 a_5\right) f_{-----}.
\end{eqnarray}
\end{minipage}}

\newpage
\section{$\psi_3$ Images of Fermion States}

The following equations give representative $\psi_3$-images of the first-generation fermion states used in the Yukawa-pairing calculations. For coloured states, the superscript $(1)$ denotes a representative colour component; the corresponding expressions for the other colour components are obtained by the same $\psi_3$ action and colour-index replacement. The final subscript labels the algebraic generation sector, so that $X_1=X$, $X_2=\psi_3(X)$, and $X_3=\psi_3^2(X)$. The formulae are written in the original Witt basis so that all trace pairings can be evaluated within a single fixed matrix realisation of $\bb{C}\ell(10)$.

\noindent\makebox[\textwidth][l]{%
\hspace*{-1.2cm}%
\begin{minipage}{1.1\textwidth}
\begin{eqnarray}
u_{L,1}^{(1)} &:=& a_1^\dagger a_4^\dagger f_{+++++}, \\
u_{L,2}^{(1)} &:=& \left(\frac{1}{4}\,a_1^\dagger a_4^\dagger+\frac{i\sqrt{3}}{4}\,a_1^\dagger\right) f_{+++++}+\left(-\frac{\sqrt{3}}{4}\,a_2 a_3 a_4^\dagger+\frac{3i}{4}\,a_2 a_3\right) f_{---++}, \\
u_{L,3}^{(1)} &:=& \left(\frac{1}{4}\,a_1^\dagger a_4^\dagger-\frac{i\sqrt{3}}{4}\,a_1^\dagger\right) f_{+++++}+\left(\frac{\sqrt{3}}{4}\,a_2 a_3 a_4^\dagger+\frac{3i}{4}\,a_2 a_3\right) f_{---++},\\
d_{R,1}^{(1)} &:=& a_1^\dagger f_{+++-+}, \\
d_{R,2}^{(1)} &:=& \left(\frac{1}{4}\,a_1^\dagger+\frac{i\sqrt{3}}{4}\,a_1^\dagger a_4\right) f_{+++-+}+\left(-\frac{\sqrt{3}}{4}\,a_2 a_3+\frac{3i}{4}\,a_2 a_3 a_4\right) f_{----+}, \\
d_{R,3}^{(1)} &:=& \left(\frac{1}{4}\,a_1^\dagger-\frac{i\sqrt{3}}{4}\,a_1^\dagger a_4\right) f_{+++-+}+\left(\frac{\sqrt{3}}{4}\,a_2 a_3+\frac{3i}{4}\,a_2 a_3 a_4\right) f_{----+},\\
\nu_{L,1} &:=& a_4^\dagger a_5^\dagger f_{+++++}, \\
\nu_{L,2} &:=& \left(\frac{1}{4}\,a_4^\dagger a_5^\dagger+\frac{\sqrt{3}}{4}\,a_1^\dagger a_2^\dagger a_3^\dagger a_4^\dagger a_5^\dagger\right) f_{+++++}+\left(-\frac{3}{4}\,a_4^\dagger a_5^\dagger+\frac{\sqrt{3}}{4}\,a_1 a_2 a_3 a_4^\dagger a_5^\dagger\right) f_{---++}, \\
\nu_{L,3} &:=& \left(\frac{1}{4}\,a_4^\dagger a_5^\dagger-\frac{\sqrt{3}}{4}\,a_1^\dagger a_2^\dagger a_3^\dagger a_4^\dagger a_5^\dagger\right) f_{+++++}+\left(-\frac{3}{4}\,a_4^\dagger a_5^\dagger-\frac{\sqrt{3}}{4}\,a_1 a_2 a_3 a_4^\dagger a_5^\dagger\right) f_{---++}, \\
e_{R,1}^{-} &:=& a_5^\dagger f_{+++-+}, \\
e_{R,2}^{-} &:=& \left(\frac{1}{4}\,a_5^\dagger+\frac{\sqrt{3}}{4}\,a_1^\dagger a_2^\dagger a_3^\dagger a_5^\dagger\right) f_{+++-+}+\left(-\frac{3}{4}\,a_5^\dagger+\frac{\sqrt{3}}{4}\,a_1 a_2 a_3 a_5^\dagger\right) f_{----+}, \\
e_{R,3}^{-} &:=& \left(\frac{1}{4}\,a_5^\dagger-\frac{\sqrt{3}}{4}\,a_1^\dagger a_2^\dagger a_3^\dagger a_5^\dagger\right) f_{+++-+}+\left(-\frac{3}{4}\,a_5^\dagger-\frac{\sqrt{3}}{4}\,a_1 a_2 a_3 a_5^\dagger\right) f_{----+}.
\end{eqnarray}
\end{minipage}}

\end{document}